\documentclass[aps,twocolumn,showpacs,floatfix]{revtex4-1}
\usepackage{graphicx}
\usepackage{amsmath}
\def\beq{\begin{equation}}
\def\eeq{\end{equation}}
\def\bea{\begin{eqnarray}}
\def\eea{\end{eqnarray}}

\begin{document}
\title{Effect of parameter mismatch on the synchronization of strongly
coupled self sustained oscillators}

\author{N. Chakrabarty}
\author{A. Jain}
\author{Nijil Lal C.K.}
\author{K. Das Gupta}\thanks{kdasgupta@phy.iitb.ac.in}
\author{P. Parmananda}
\affiliation{Department of Physics,
Indian Institute of Technology Bombay,
 Mumbai-400076, India.}

\date{\today}


\begin{abstract}
In this paper we present an experimental setup and an associated mathematical model to study the synchronization of two self sustained strongly coupled mechanical oscillators (metronomes). The effects of a small detuning in the internal parameters, namely damping and frequency, have been studied. Our experimental system is a pair of spring wound mechanical metronomes, coupled by placing them on a common base, free to move along a horizontal direction.
In our system the mass of the oscillating pendula form a significant fraction of the total mass of the system, leading to strong coupling of the oscillators.
We modified the internal mechanism of the spring-wound "clockwork" slightly, such that the natural frequency and the internal damping could be independently tuned. Stable synchronized and anti-synchronized states were observed as the difference in the parameters was varied. We designed a photodiode array based non-contact, non-magnetic position detection system driven by a microcontroller to record the instantaneous angular displacement of each oscillator and the small linear displacement of the base coupling the two. Our results indicate that such a system can be made to stabilize in both in-phase anti-phase synchronized state by tuning the parameter mismatch. Results from both numerical simulations and experimental observations are in qualitative agreement and are both reported in the present work.
\end{abstract}

\pacs{05.45.Xt, 05.65.+b, 89.75.-k}

\maketitle

\section{Introduction}
Synchronization is a phenomenon wherein interacting, oscillating systems adjust a property of their motion. Such phenomena are abundant in physical, biological as well as social systems \cite{Mirollo,Glass,Thornburg,Rivera,Boccaletti}. Huygens, known for his work on the wave nature of light, was probably the first to observe and document this phenomenon. He reported observing two pendulum clocks, suspended from a common support, to synchronize in {\it anti-phase}.  The phenomenon re-established itself even if disturbed by some external interference. He correctly ascribed the reason for this to the minute mechanical oscillations travelling through the supporting beam \cite{Pikovsky, Huygens}.
In these experiments, the mass of the pendula formed a very small fraction of the total mass of the system (estimated
to be less than $\sim{1\%}$), compared to our experiments reported here (at least $\sim{25\%}$).\\

A metronome is a non linear self sustained oscillator designed for timing musical exercises - its internal mechanisms can be accessed and modified  with relative ease, making it an attractive model system. The non-linearity in the motion of a single metronome arises from two principal sources. First, the amplitude of the pendulum of the oscillator is fairly large (reaching 20-30 degrees on each side). Second and more importantly, the mechanism that sustains the motion of the pendulum and compensates for frictional losses, injects some energy in small pulses.  We will show that modelling this inherently non-linear injection mechanism correctly is important in understanding the asymptotic states. The phase space of a single metronome is two dimensional, and hence asymptotically only steady states or periodic oscillations are possible. However  synchronization phenomena can be observed when two or more metronomes are coupled \cite{Ulrichs}. Such coupled  ${\lq\lq}$clocks" show a variety of collective behaviour and continue to receive  attention \cite{Bennet,Pantaleone,K.Czolczynski, Czol2, Czol3,qianghu}, since they serve as a model of complex dynamics in coupled systems. Similar mechanical clocks have also been used to study collective behaviour of an $N$-oscillator system (the Kuramoto transition)\cite{Boda-Kuramoto}.

Pantaleone \cite{Pantaleone}  introduced an experimental setup and mathematical model of two (or more) metronomes  coupled by placing them on a horizontally movable base. He reported observing primarily in-phase synchronized states of the coupled metronomes. More recently,  Czolczynski {\it et al} \cite{ K.Czolczynski, Czol2, Czol3} have studied a qualitatively similar problem of synchronization in coupled clocks which replicates the original Huygens' experiment. Wu et al. \cite{Wu} have studied the effect of base damping and initial conditions on the asymptotic state of synchronization and reported on the occurrence probability of different states of synchronization as a function of the base damping. A survey of the existing literature shows that  {\it in-phase} and {\it anti-phase} synchronization, as well as $\lq\lq$oscillator death" regimes  have been reported in the coupled clock system.  Our paper focuses on the effect of parameter (frequency and internal damping) mismatch on the state of synchronization of the coupled metronomes. We show that it is possible for the same pair of oscillators to have an asymptotically stable in-phase state or an anti-phase state depending on the
{\it mismatch} of some experimentally tunable internal parameter.  \\

We couple two metronomes, optionally made non-identical by introducing a small difference in their internal damping parameter or frequency and study the  long time behaviour. The metronomes are bi-directionally coupled by placing them on a horizontally movable base. We designed an opto-electronic position tracking system using  linear photo sensor arrays to track the motion of the individual metronomes and the movable platform connecting them simultaneously. For our mathematical model we follow equations of motion similar to the ones  described in \cite{Pantaleone, Ulrichs, K.Czolczynski} with a modification incorporating the action of the escapement mechanism. It has been customary to model the non-linear oscillators  using the Van der Pol (or Rayleigh) equation \cite{Pantaleone,VDP}. However this form of the driving equations ({\it e.g.} as in ref \cite{Pantaleone}) does not allow independent tuning of the energizing and dissipative terms. We show that it is necessary to modify the equation to account for observed motions. The energy put in by the escapement mechanism is modeled as an impulse that acts only when the angular displacement is within a certain small range and the velocity is directed away from the resting (zero) position. This impulse or $\lq\lq$kick" is the key non-linear element of the problem. Our model equations admit stable limit cycles and separates out the dissipation and the impulse delivered by the escapement mechanism in each cycle.\\

The paper is organized as follows: Section \ref{sec:II} describes the experimental setup and the data acquisition mechanism, Section \ref{sec:III} describes the experimental results. In Sec. \ref{sec:IV} we discuss the mathematical model and in Sec. \ref{sec:V} the numerical results. Finally we summarize our work in Sec. \ref{sec:VI}.


\begin{figure*}[t]
\begin{center}
\includegraphics[width=\textwidth, bb= 0 0 1710 933]{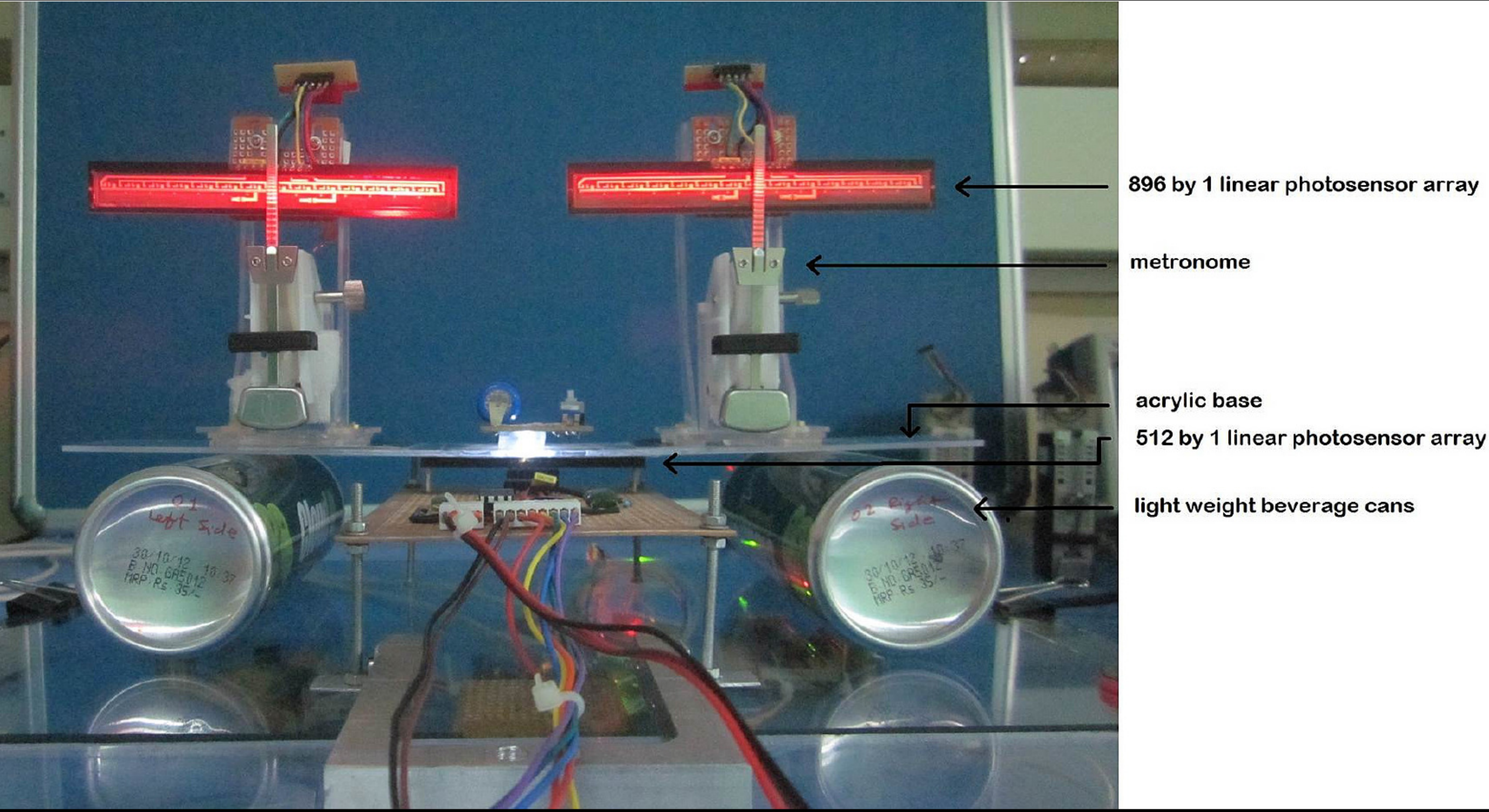}
\end{center}
\caption{(Color online) Experimental setup showing the two metronomes coupled by placing them on a horizontally movable base. Two 896$\times$1 linear photosensor array for each metronome and one 512$\times$1 linear photosensor array are used to determine the instantaneous positions. The
photosensor arrays are driven by microcontrollers which convert the light intensity distribution data to instantaneous position, once in less
than 2 milliseconds.}
\label{setup-photo}
\end{figure*}

\section{Experimental Setup and Data Acquisition}
\label{sec:II}

The experimental setup (Fig.\ref{setup-photo}) consists of two metronomes \cite{Wittner} mounted on a common base that rests on two light, hollow cylinders (beverage cans). The $\lq\lq$escapement mechanism" \cite{Escapement} delivers pulses of energy from a coiled spring (twice every cycle) and keeps the oscillators moving, compensating for  frictional losses. The oscillations are sustained till the spring is fully uncoiled. In our experiment this takes about 20 minutes.\\

The lightweight acrylic base (50 gram) and the cylinders (10 grams each) ensure horizontal motion of the base with minimum dissipation. To further reduce the weight, the metronome's  mechanisms were removed from their plastic bodies and were fitted to customized acrylic housings. The motion of the metronome hand is tracked by an opto-electronic position tracking system which traces the $\lq\lq$center of mass" of the  shadow of the oscillating pointer of the metronome, on a linear photo sensor array \cite{Photosensor} placed behind it. The sensor has 896 pixels uniformly distributed over a line  112 millimeters in length. The moving hand is illuminated by a strip of light shaped by a cylindrical lens of appropriate focal length. The separation between the plane of the moving hand and the photosensor array is approximately 1mm and the shadow is  sharply defined.  Our way of mounting the sensing mechanism ensures that the displacement of each oscillator is measured in its own reference frame. The method is also efficient since it is non-contact and non-magnetic, thus introducing  no additional friction or stray eddy currents that may be associated with the magnetic detection mechanism. A  similar arrangement is made to detect the much smaller motion of the
base (typically 1-2 mm), using a narrow slit and a 512$\times$1 photosensor array with a finer pitch. \\

The tracking system is driven by one microcontroller \cite{Arduino} for each sensor array.  The controllers are programmed to scan each photodiode array fully in 2 milliseconds and generate an output voltage proportional to the pixel number of the $\lq\lq$center of mass" of the shadow. All output voltages are then read by a data acquisition system 50 times a second and stored in a computer.  The full details of the optics, electronics and the code used to run the microcontroller will be given elsewhere.

\begin{figure}[h!t]
\includegraphics[width=0.45\textwidth, clip= ]{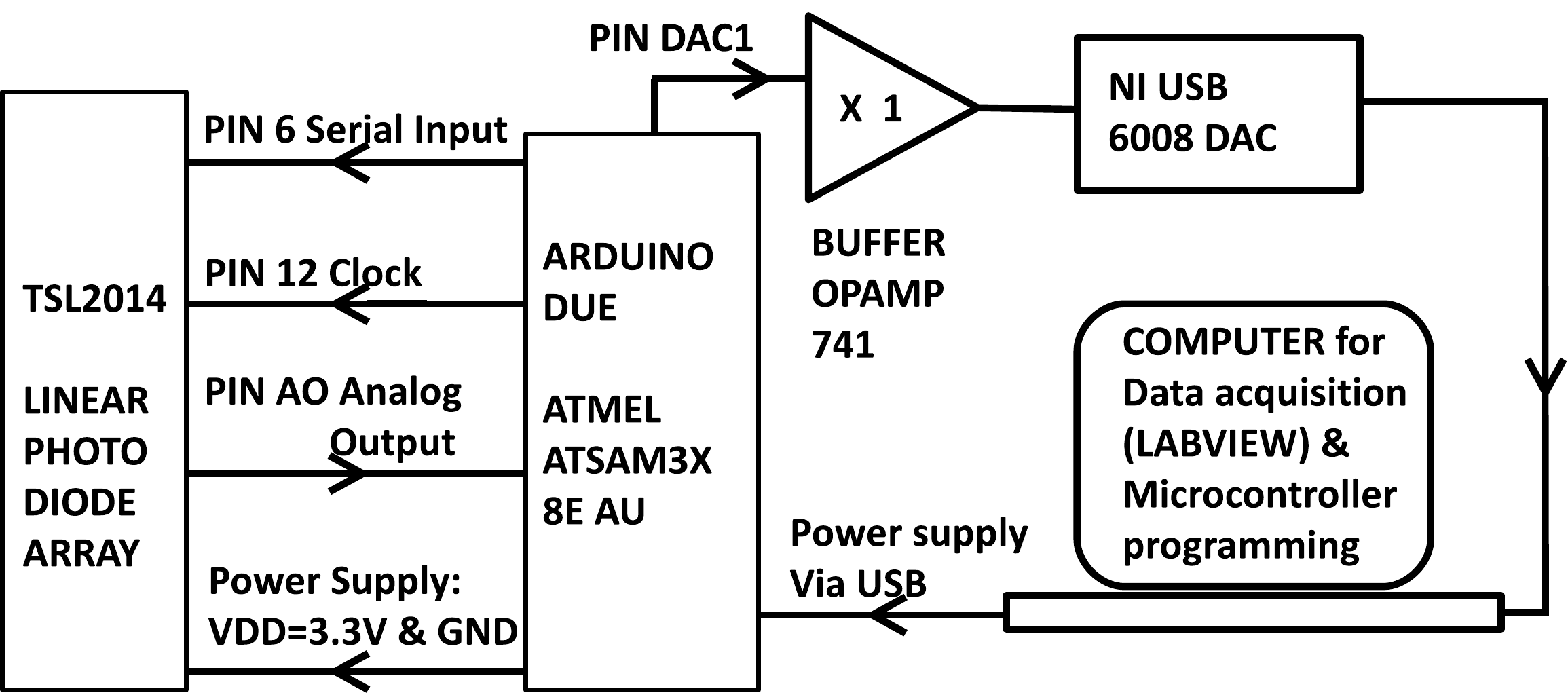}
\caption{Schematic for acquiring the data of the instantaneous position of the metronome from the photosensor array via the ARDUINO Due Board \cite{Arduino} to the computer.}
\label{2}
\end{figure}

Figure \ref{2} shows the flowchart for acquiring the data  from the photosensor array via the ARDUINO Due  microcontroller board \cite{Arduino} to the computer.

\section{Experimental Results}
\label{sec:III}

\begin{figure}[t]
\begin{center}
\includegraphics[width=0.45\textwidth,clip=]{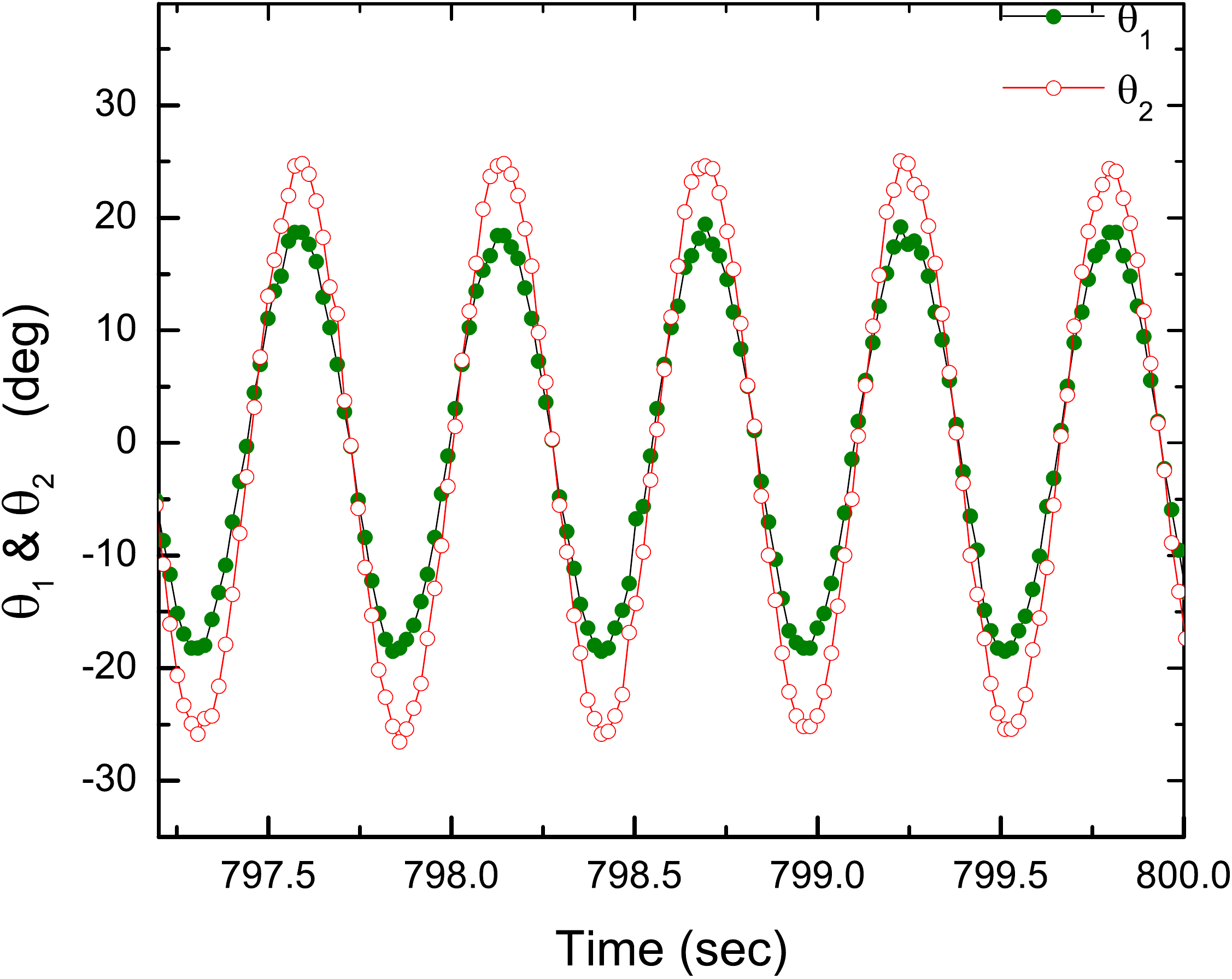}\\
\includegraphics[width=0.45\textwidth,clip=]{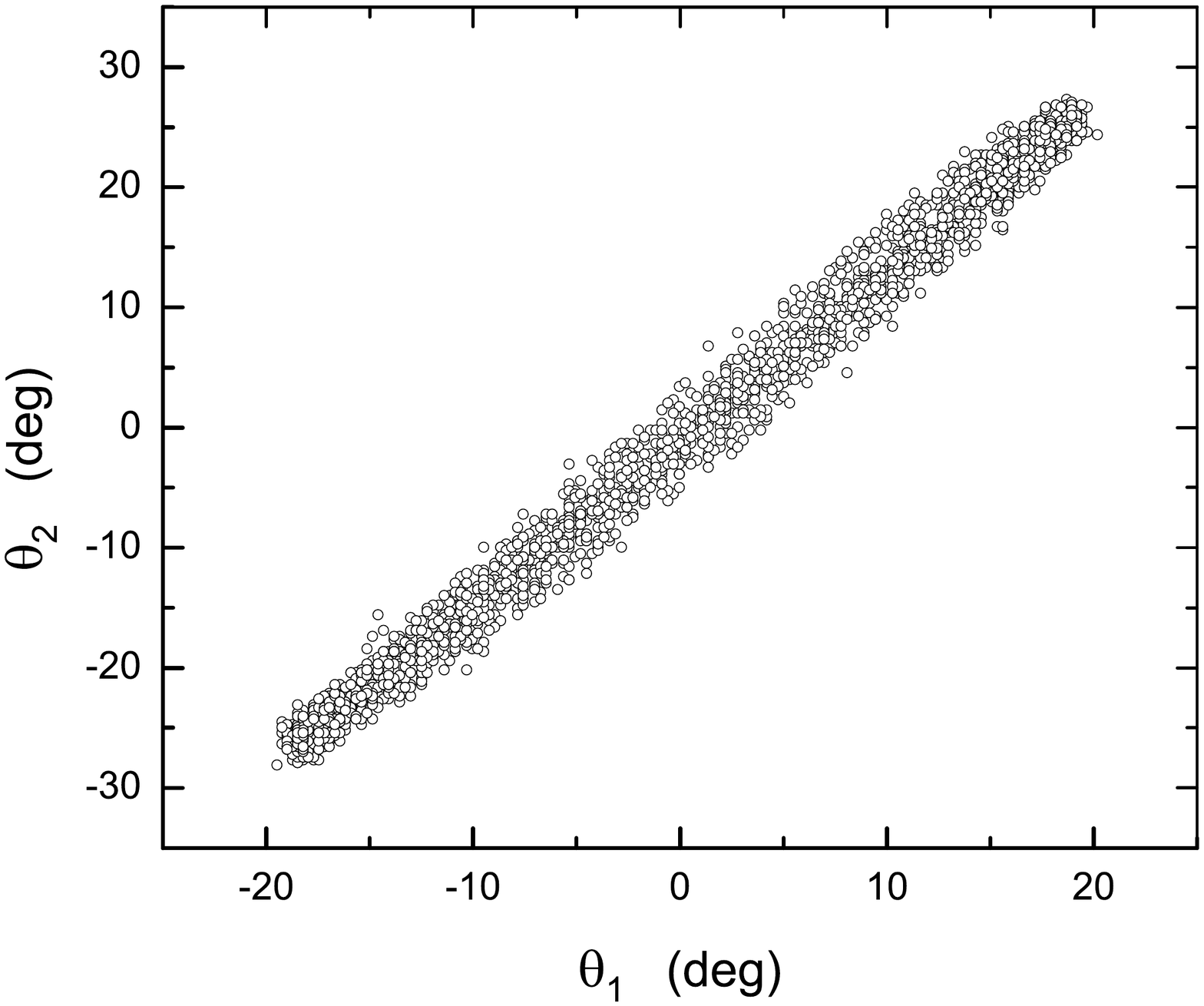}
\end{center}
\caption{(Color online) Experimental Results showing the final in-phase synchronized states of two metronomes. The upper panel figure shows the timeseries of $\theta_1$,$\theta_2$ and the lower panel figure shows the plot $\theta_2$ vs. $\theta_1$. }
\label{3}
\end{figure}

In the experiments, when the two metronomes were coupled, with their parameters matched as closely as possible, they were found to  oscillate with synchronized phases as in ref\cite{Pantaleone}. Figure \ref{3} shows the time series of $\theta_1$ and $\theta_2$ and the plot for $\theta_1$ vs. $\theta_2$ which corresponds to complete in-phase synchronization.

\begin{figure}[t]
\begin{tabular}{c}
\includegraphics[width=0.45\textwidth,clip=]{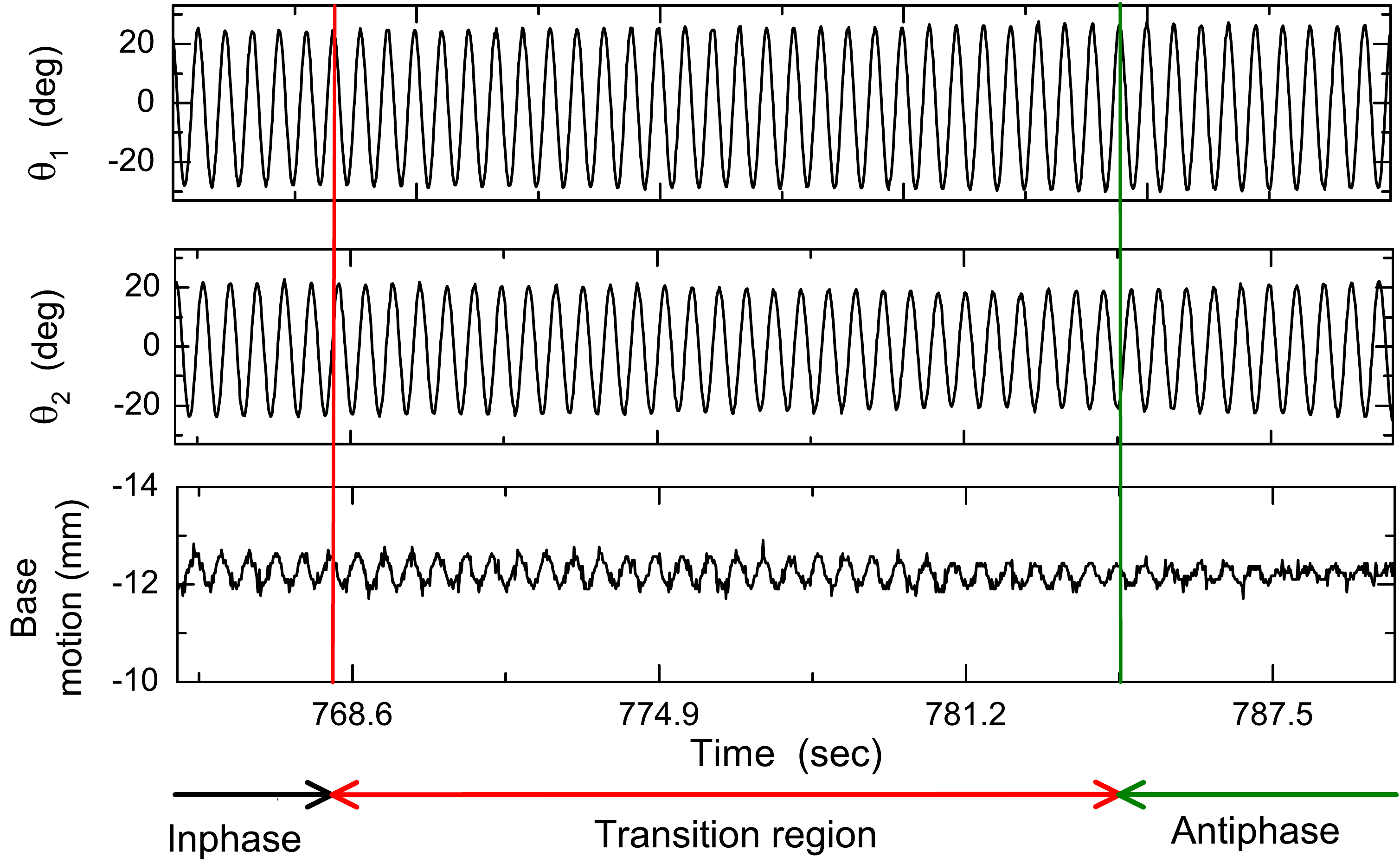}\\
\includegraphics[width=0.45\textwidth,clip=]{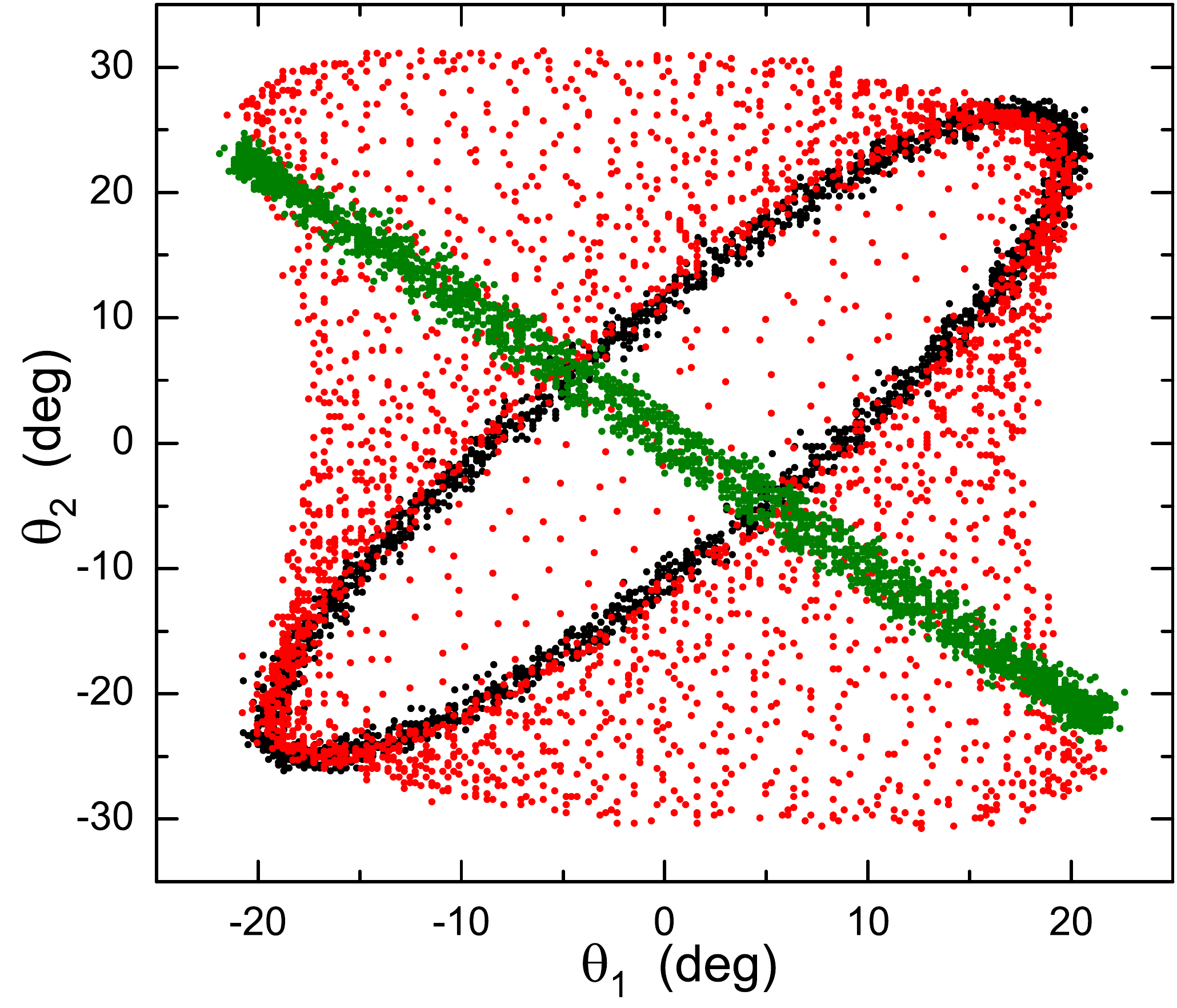}\\
\end{tabular}
\caption{(Color) Top:Experimental Results showing the time series and the phase plot for the transition from an initial in-phase condition to the asymptotic anti-phase synchronized state. (Bottom) In the phase plot the black region corresponds to the initial in-phase state, the red region corresponds to the transition regime and the green region corresponds to the final anti-phase synchronized state. The time series for the oscillations of the base is also plotted, the decay of which confirms the onset of the anti-phase synchronized state. }
\label{4}
\end{figure}

With sufficient damping in one of the metronomes, introduced by pressurizing the spring using a screw, the system synchronizes in anti-phase state, even if started from an initial in-phase condition. This appears to be  the only asymptotically stable state for higher values of parameter mismatch (Fig. \ref{4}). The anti-phase state is
reached from all initial conditions sufficient to start the oscillation - since the escapement mechanism of a
clock engages only if the displacement exceeds a certain critical value, the initial conditions require that $\theta_1(t=0)$ and $\theta_2(t=0)$ be larger than this. In our case this was approximately 15 degrees.
The mass of each pendulum ($m$)is approximately $m_1=m_2=28 {\rm gms}$ and the total weight of each metronome, taking into account
the acrylic housing, photodiode array etc is 94 gms. The base coupling the two oscillators weighs 50 gms and the
aluminium can rollers 10gms each. If $M+2m$ is taken as the total mass of the system, we have  $2m/(M+2m) {\approx}0.24$. It is worth
recalling that this ratio in the original data reported by Huygens and subsequent reproductions of the experiment
as reported in ref \cite{Bennet} is  about two orders of magnitude smaller. Our experiments  thus pertain to
a regime where the dynamics can have significant differences.

\begin{figure*}[t]
\begin{center}
\includegraphics[height=5.8cm,clip=]{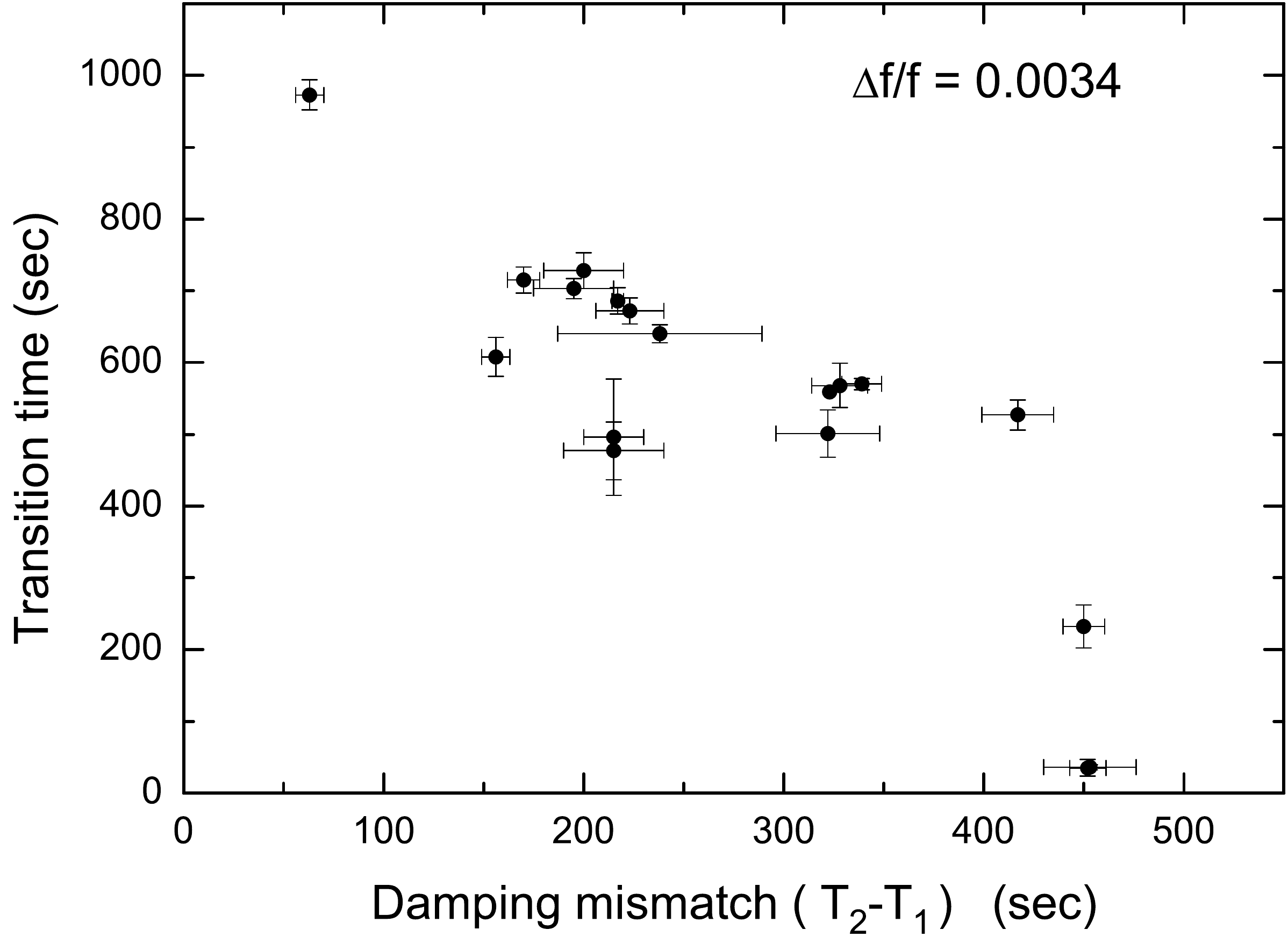}\hspace*{0.5cm}\includegraphics[height=5.8cm,clip=]{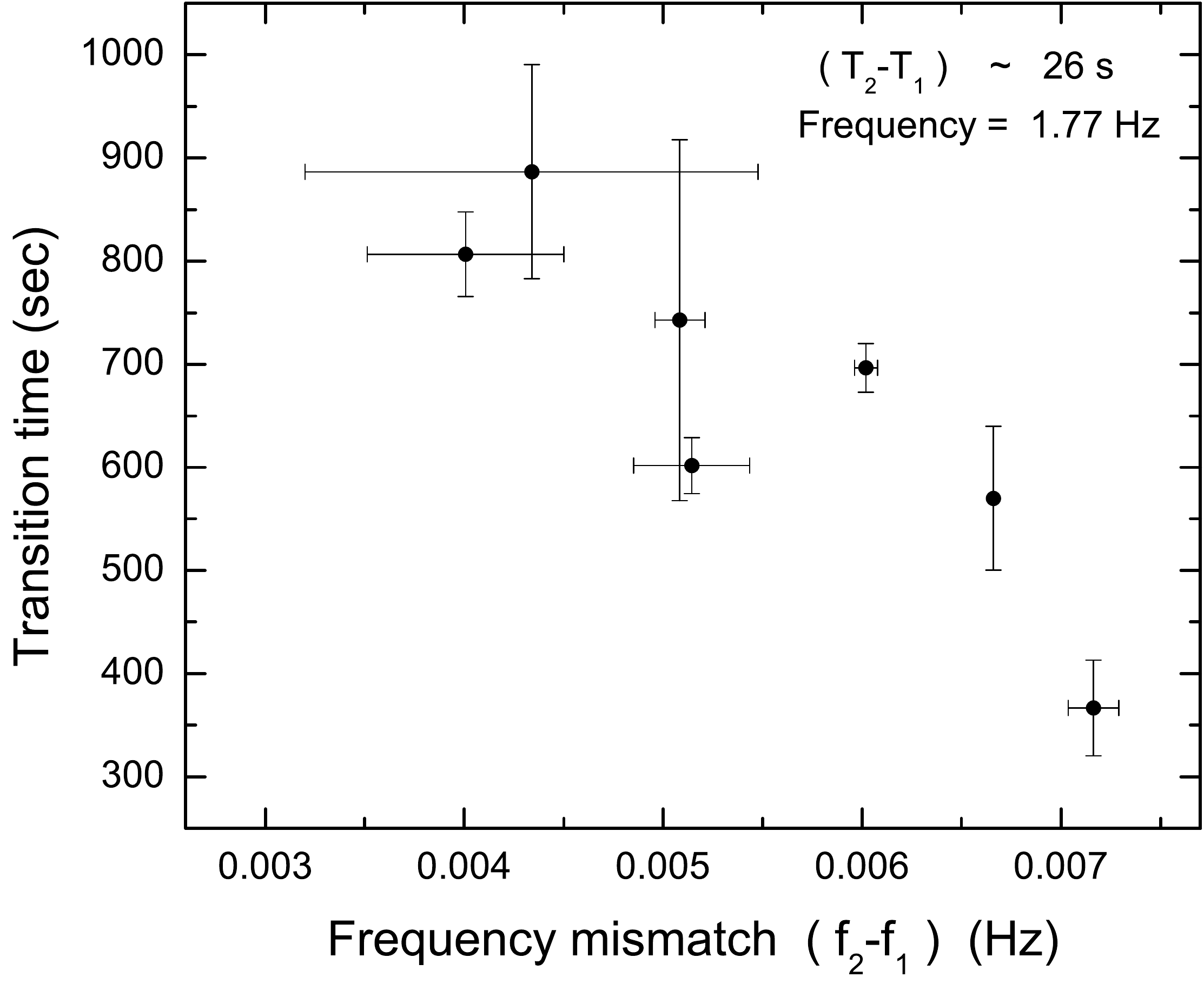}
\end{center}
\caption{(Left)Experimental results for the time taken to settle into the anti-phase synchronized state from an initial in-phase state as a function of the damping mismatch. Here the frequency mismatch was measured to be fixed at 0.34\%. (Right)Experimental results for the time taken to settle into the anti-phase synchronized state from an initial in-phase state as a function of the frequency mismatch. Here the damping mismatch $\dfrac{T_2 - T_1}{T_2}$ was measured to be fixed at 2.17\%.}
\label{5and6}
\end{figure*}

Since the relative damping between the two metronomes is difficult to determine explicitly, we take the lifetime of each oscillator (quantified as the time taken to completely die out, $T_1$ and $T_2$) as a measure of
the damping and the difference $(T_2 - T_1)$ as the relative damping. We measure the frequency mismatch to be very small, about $0.34\%$ of their uncoupled frequencies. In each measurement, we measure $T_1$ and $T_2$ and the time taken to settle to the final anti-phase synchronized state starting from an initially in-phase condition. The results are shown in Fig. \ref{5and6}(left).

We obtain similar results when a small frequency mismatch is introduced in the system, by varying the mass of one of the metronome bobs. The frequency mismatch is determined by recording the beat frequency of the two metronomes when they are uncoupled. For each measurement the time taken to settle into the asymptotic anti-phase synchronized state was taken for a given frequency mismatch $\dfrac{f_2-f_1}{f_1}$ shown in Fig. \ref{5and6}(right).

\section{Mathematical Model}
\label{sec:IV}

\begin{figure}[h]
\begin{center}
\includegraphics[width=0.45\textwidth, clip=]{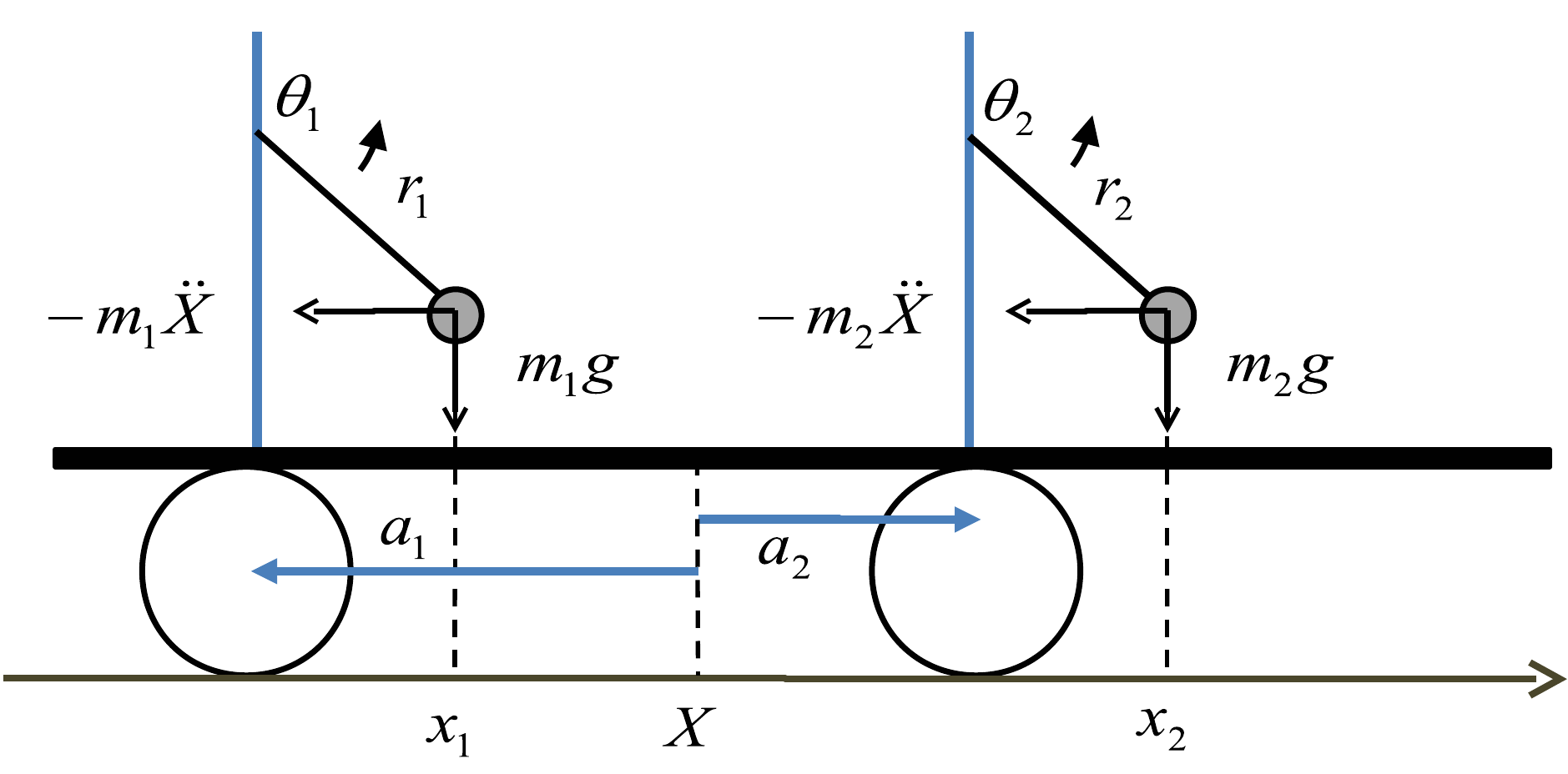}
\end{center}
\caption{Diagram showing the forces acting on each metronome bob and the variables defined in our equations.}
\label{7}	
\end{figure}

Figure \ref{7} shows the forces acting on each of the metronome bobs. The plane of oscillation of the metronomes' bob is perpendicular to the plane of oscillation of the base. Following the formulation of \cite{Pantaleone} we proceed to derive the equations of motion in an n-metronome system.
The center of the mass of the metronomes + base system is at:
\beq
x_{cm} = \frac{MX + \sum\limits_{i=1}^{n}  m_ix_i  }{M +  \sum\limits_{i=1}^{n}  m_i }
\eeq
where $M$ is the mass of the base, $X$ is the position of the center of mass of the base, $m_i$ the mass, and $x_i$ the position of the bob of the \textit{i}-th metronome.
\beq
x_i = X + a_i + r_{i}\sin\theta_i
\eeq

Here $a_i$ is the constant position difference of the equilibrium position of the  \textit{i}-th metronome from the center of mass of the
 system without the pendula and $r_{i}$ is the distance of the \textit{i}-th metronome's center of mass from its pivot point. For our purposes we can consider:
\beq
\frac{d^2x_{cm}}{dt^2}=0
\label{eqn-com-does-not-move}
\eeq

We have $n+1$ variables - $X$ and the $n$ angular variables - $\theta_i$, but there is one constraint given by equation \ref{eqn-com-does-not-move}. We therefore need $n$ independent equations.

The equation of constraint (eqn.\ref{eqn-com-does-not-move}) gives

\beq
\frac{d^2{X}}{dt^2} = -\frac{1}{M + \sum\limits_{i=1}^{n}m_i}\frac{d^2{}}{dt^2}\left( \sum\limits_{i=1}^{n}m_{i}r_{i}\sin\theta_{i}\right)
\eeq

Since the angular displacements are measured in a frame fixed on an accelerating base,
the gravitational force, the internal forces and the pseudo force  contribute to a torque on the $i^{th}$ pendulum calculated about
its pivot point.
Therefore the coupled equation of motion of the \textit{i}-th metronome in an n-metronome system can be written as \cite{Pantaleone}:
\begin{widetext}
\begin{equation}
\frac{d^2\theta_i}{dt^2} = -\omega_i^2\sin\theta_i -\frac{r_i}{I_i}F_{i,internal}(\theta_i,\dot{\theta_i})
 +\beta_i \cos\theta_i\frac{d^2}{dt^2}\sum\limits_{j=1}^n m_{j}r_{j}\sin\theta_j
\label{eqn-of-motion-of-ith-metronome-bob}
\end{equation}
\end{widetext}
Here
\beq
\omega_i^2 = \frac{m_{i}r_{i}g}{I_i}
\eeq

is the square of the angular frequency of the undamped, uncoupled small amplitude oscillator, $r_{i}$ is the distance  of the metronome bob from the pivot point and $I_i=m_{i}{r_{i}}^{2}$ is the moment of inertia of the bob about the axis perpendicular to the plane of oscillation of the  metronomes.

\beq
\beta_i= \frac{m_ir_{i}}{I_i}\frac{1}{M+\sum\limits_{j=1}^{n}m_j}
\eeq

is a measure of the coupling strength. If all $m_i$ and $r_i$ are equal, then only they can be pulled out of the summation and the
coupling coefficient can be refined such that it becomes dimensionless.

$F_{i,internal}(\theta,\dot{\theta})$ models the escapement mechanism of the metronome which sustains the oscillations.\\

The terms in Eq.(5) represents the following: The first term is the angular acceleration, the second term is the torque due to the gravitational force. The last term represents the torque due to the pseudo force:$F_{i,pseudo} = -m_i \dfrac{d^2{X}}{dt^2}$ which we have to take into account since the metronome is in a non-inertial frame of reference.\\

The third term $ F_{i,internal}(\theta,\dot{\theta})$ models the escapement mechanism of the metronome \cite{Escapement} and takes into account both the damping and the impulse which the unwinding spring "feeds" into the oscillator. Since the rolling cylinders are light and provide a low friction, we neglect their effect on the equations of motion and assume the base friction to be negligible. We assume $F(\theta,\dot{\theta})$ to be of the following form:

\beq
  F_{i,internal}(\theta_i,\dot{\theta_i}) = \mu_{i} \theta_{i}^2\frac{d\theta_{i}}{dt} - f_{i}(\theta_{i},\dot{\theta_{i}})
\eeq

where:

\beq
 f_{i}(\theta_{i},\dot{\theta_{i}}) =
\left\{
	\begin{array}{rlrcr}
		c_{i}\dot{\theta_{i}}   & \rm{ if } & \theta_0-\delta\theta_0 &< \theta_{i}< & \theta_0+\delta\theta_0\\
&& \mbox{ \& } \dot{\theta_{i}} > 0&&\\
		- c_{i}\dot{\theta_{i}} & \rm{ if}&  -\theta_0-\delta\theta_0 &< \theta_{i}< & -\theta_0+\delta\theta_0\\
&& \mbox{ \& } \dot{\theta_{i}} < 0&&\\

                      0  & &\rm{otherwise} &&
	\end{array}
\right.
\eeq

  $F_{i,internal}(\theta_{i},\dot{\theta_{i}})$ is separated into damping and forcing components and therefore allows independent control of the individual components. Since in the experiments we could control the damping of the metronomes without affecting the forcing, we need to formulate $F_{i,internal}(\theta_{i},\dot{\theta_{i}})$ in a way that allows for independent control of each component. \\

In the previous works on the subject \cite{Pantaleone, Ulrichs} a Van der Pol (or Rayleigh) term \cite {VDP} of the form:
\beq
F(\theta,\dot{\theta})= \mu \left(\left(\frac{\theta}{\theta_0}\right)^2 -1 \right) \frac{d\theta}{dt}
\eeq
has been used.\\

  This form of the force therefore decelerates the motion for $\theta > \theta_0$ and accelerates it for $\theta < \theta_0$ and leads to limit cycles which are typical of Van der Pol oscillators \cite{VDP}. However since this term takes into account both the forcing and the damping of the metronome, it does not allow independent control of the two parameters.

Thus $f_{i,internal}(\theta_i,\dot{\theta_i})$ is a constant impulse imparting force which acts only when the metronome bob is in a certain angular range, twice during each of the metronome's period of oscillation and models the $\lq\lq$kick" force the bob receives from the wound up spring. The results remain qualitatively invariant by using a constant energy imparting force or by adding an additive Gaussian white noise to $ f_{i,internal}(\theta_i,\dot{\theta_i})$. A somewhat similar form is used by K. Czolczynski et al. \cite{K.Czolczynski,Czol2} for their model of coupled clocks.

\begin{figure}[h]
\begin{center}
\includegraphics[width=0.5\textwidth,clip= ]{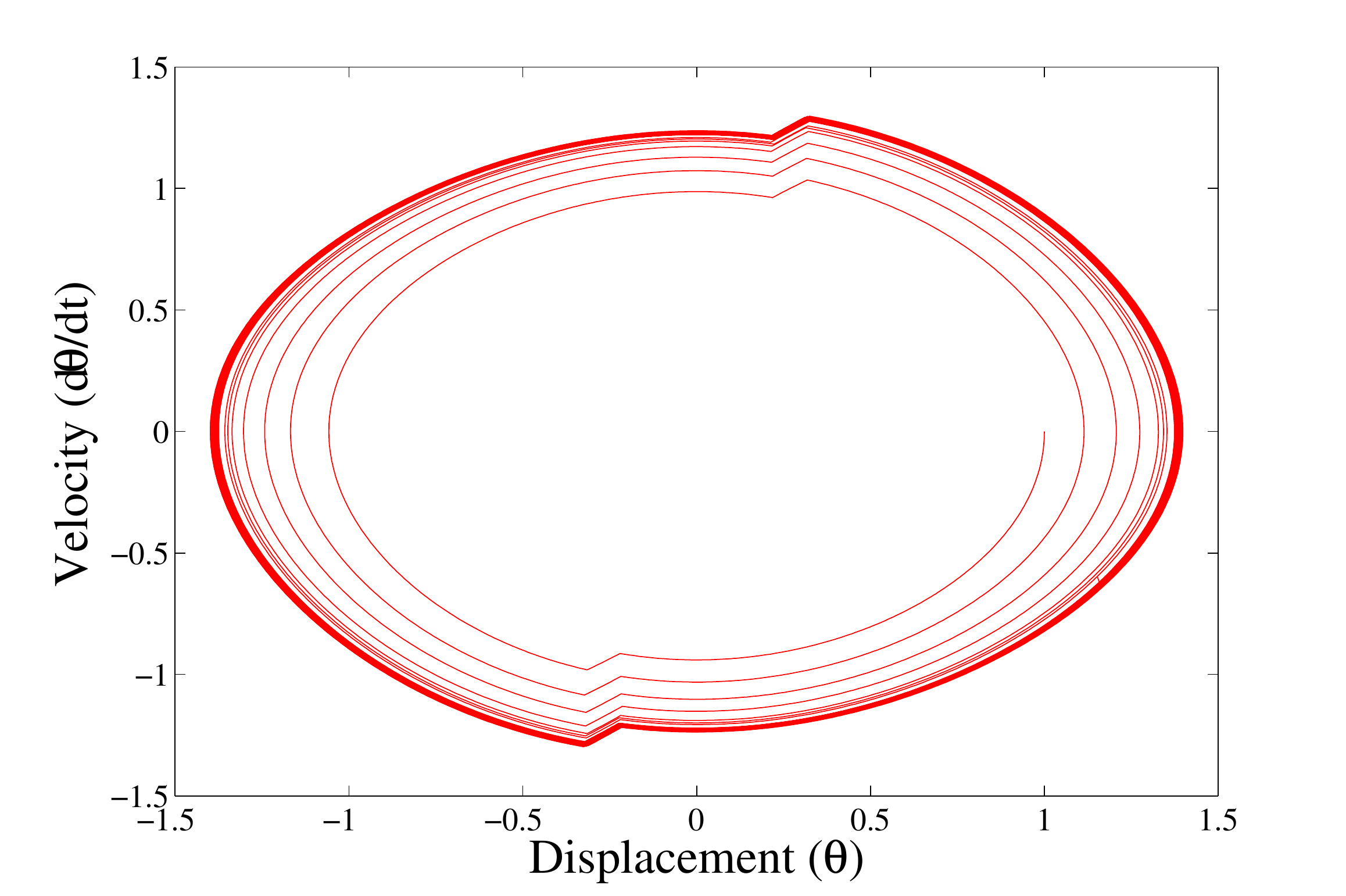}
\end{center}
\caption{Numerical results showing the limit cycle for a single metronome. The small distortion visible in the plot signifies the action of the impulse $f(\theta,\dot{\theta}).$}
\label{8}
\end{figure}

\begin{figure}[h]
\includegraphics[width=0.5\textwidth, height=6cm]{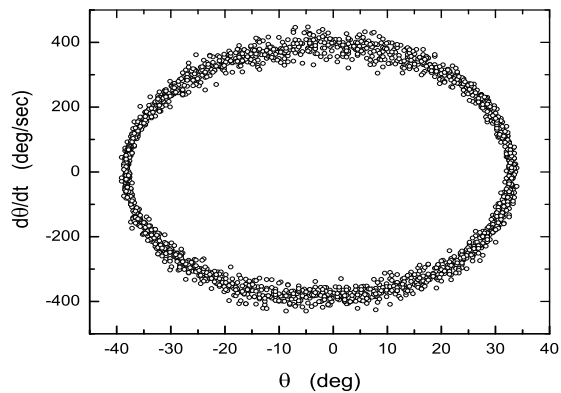}
\caption{Experimental results showing the $\theta$ vs. $\dot{\theta}$ plot for a single metronome placed on a moving base.
\label{9}}
\end{figure}

The phase plot of a single metronome(limit cycle), kept on a movable base, obtained by numerically integrating Eq.(5) using a standard integration method (like fourth order Runge-Kutta) is shown in Fig.( \ref{8}) by taking values of the parameters as:$\mu=0.1$ , $c= 1$, $\theta_0 = 0.26$, $\delta\theta_0 = 0.05$. $\beta = 0.01$. The small glitches visible in Fig.\ref{8} come at the points where the escapement mechanism delivers the impulse to keep the system running. It is possible to see this glitch experimentally if the angular displacement data is taken at a much higher rate.
The rate (50 pts/sec) we collected the data is not fast enough for this glitch to be seen. However by taking high speed movies of the oscillating pointer we were able to spot this. The parameter values are obtained from experimental data and are scaled by taking $\omega = 1$. The corresponding experimental plot is shown in Fig.(\ref{9})

\vspace{-1cm}

\section{Numerical Results}
\label{sec:V}
\begin{figure}[h]
\includegraphics[width=0.5\textwidth,clip=]{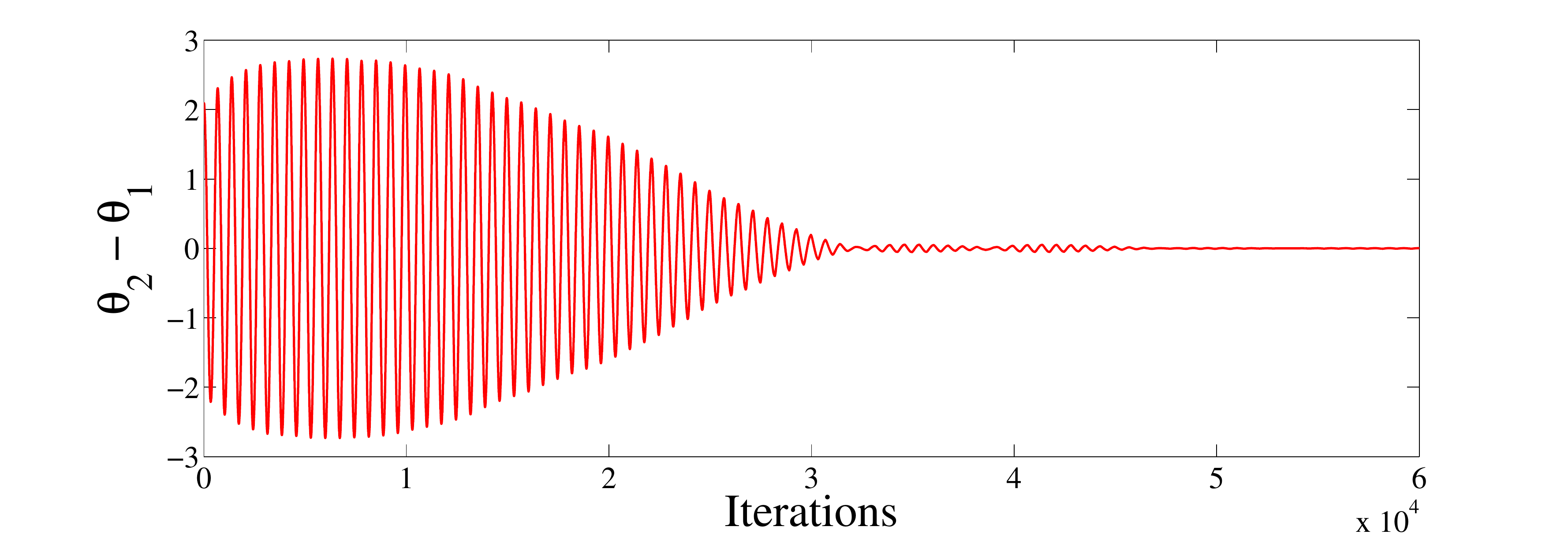}
\caption{Numerical results showing the error function ($\theta_2 - \theta_1)$ decaying to zero, therefore signifying complete in-phase synchronization, for two identical metronomes.}
\label{10}
\end{figure}

We numerically integrate Eq.(5) for \textit{i}=2 signifying a two metronome system on a moving base.
As is demonstrated in \cite{Pantaleone, Ulrichs} the two metronomes is seen to synchronize in-phase for parameter matched conditions.The parameter values are taken as: $\omega_1=\omega_2=1.0$, $\mu_1=\mu_2=0.1$ , $c_1= c_2= 1$, $\theta_0 = 0.26$, $\delta\theta_0 = 0.05$. $\beta = 0.01$. It is to be noted that rather than the absolute values of the parameters their relative values is the important factor and as long as the parameter values are matched, the metronomes are seen to synchronize in phase. Figure \ref{10} shows the error function $(\theta_2-\theta_1)$ as a function of time, which decays to 0, denoting in-phase synchronization.

\begin{figure*}[ht!]
\includegraphics[width=\textwidth,clip=]{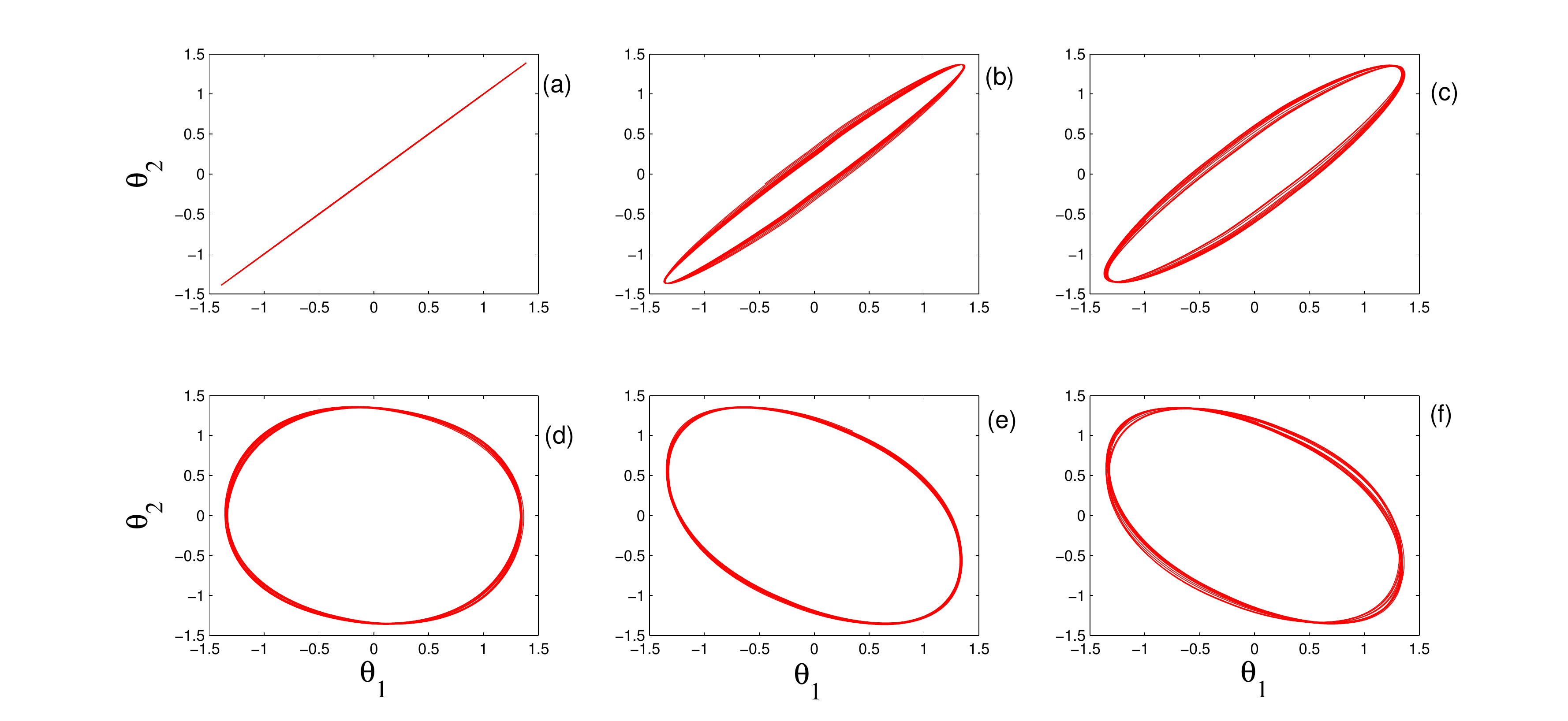}
\caption{Numerical results of $\theta_1$ vs. $\theta_2$ in the asymptotic state, starting from a random initial phase difference, for different values of the damping parameter mismatch (denoted as $\dfrac{\mu_2 -\mu_1}{\mu_1}$. Mismatch= (a) 0\%, (b) 5\%, (c) 10\%, (d) 15\%, (e) 20\%, (f) 23\%. For generating the plots the initial conditions chosen were: $\theta_1=1.0$, $\dot{\theta_1}=0$, $\theta_2=-1.1$, $\dot{\theta_2 }=0$.}
\label{13}
\end{figure*}

If we mismatch the damping parameters of the metronomes, the are seen to synchronize with a constant phase difference. This is evident from Fig. \ref{13} which shows asymptotic state of $\theta_1$ vs. $\theta_2$  for different values of mismatch. Here $\mu_2$ is varied, keeping $\mu_1=0.1$ constant. If the steady state dynamics of $\theta_1$ and $\theta_2$ are approximately sinusoidal, having a common frequency
and a constant phase difference, then:

\beq
\theta_1 = A\cos(\omega t) ~\mbox{and}~ \theta_2=B\cos(\omega t + \phi) \nonumber\\
\eeq
which implies:
\beq
\frac{\theta_1^2}{A^2} + \frac{\theta_2^2}{B^2} - \frac{2\theta_1\theta_2 \cos\phi}{AB} = \sin^2\phi
\eeq

That is, if in their asymptotic states $\theta_1$ vs. $\theta_2$  traces out an ellipse, we conclude that they are in a phase locked state.
It is to be noted that the $\theta_1$ vs. $\theta_2$  ellipse is not smooth due to oscillations about the limit cycle. Also a change in the initial conditions, leads to small variations in the final phase locked state. Therefore the resultant ellipse was fitted using a standard curve fitting routine so as to extract the phase difference of the two oscillators directly from the plot. We use the Taubin algorithm \cite{Taubin, Taubin2} for this purpose and extract the phase from the fitted ellipse (details in the Appendix section). Figure \ref{11} shows the dependence of the constant phase difference as a function of the damping parameter mismatch (in \%) denoted as $\dfrac{\mu_2 - \mu_1}{\mu_1}$. The data has been averaged over different sets of initial conditions.

\begin{figure}[ht!]
\includegraphics[width=0.5\textwidth,clip=]{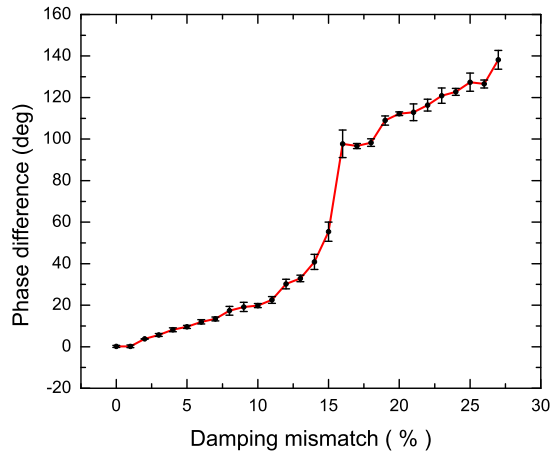}
\caption{Numerical results showing the relationship between the constant phase difference in the asymptotic state and the damping parameter mismatch $\bigg(\dfrac{\mu_2 - \mu_1}{\mu_1}\bigg)$.}
\label{11}
\end{figure}

From Fig. \ref{11} it is seen that the constant phase difference grows as the damping mismatch is increased, reaching near anti-phase conditions. After about 27\% damping mismatch, no clear relationship between $\theta_2$ and $\theta_1$ exists and the metronomes move away from a synchronized state.\\

\begin{figure}[h]
\includegraphics[width=0.5\textwidth,clip=]{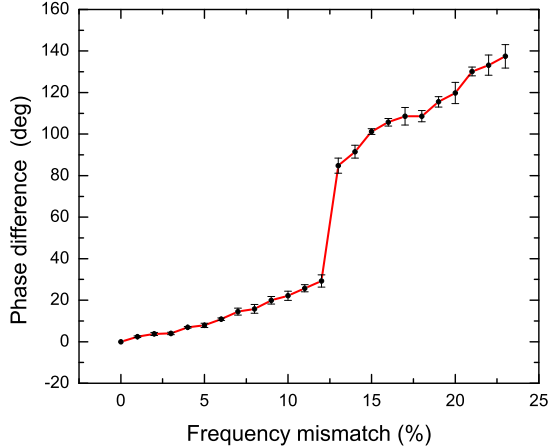}
\caption{Numerical results showing the relationship between constant phase difference in the asymptotic state and the uncoupled frequency mismatch $\bigg(\dfrac{\omega_2 - \omega_1}{\omega_1}\bigg)$.}
\label{12}
\end{figure}

Similar results are obtained for a mismatch in the undamped,uncoupled frequency parameter given by Eq.(6) by setting $\omega_1=1.0$ and varying $\omega_2$. It is a well documented fact \cite{Pikovsky} in the study of synchronization that if oscillators with a difference in their uncouled frequencies are coupled, till a small difference, they adjust their individual frequencies to a common value. In our system of coupled metronomes having a difference in their uncoupled frequencies, we observe the same behaviour. The frequency of the individual oscillators adjust to a common value and $\theta_1$ vs. $\theta_2$ synchronizes in a phase locked state. As before, we use the Taubin algorithm \cite{Taubin, Taubin2} to extract the phase difference. Figure \ref{12}shows the constant phase difference as a function of the frequency mismatch (in \%) denoted as: $\dfrac{\omega_2 - \omega_1}{\omega_1}$. From this figure, it is evident that the metronomes synchronize near in-phase condition for small difference in the uncoupled frequencies, which grows out to near anti-phase condition (about 140 degrees) for higher mismatch in the frequencies.

Therefore the constant phase difference grows as the frquency mismatch is increased, reaching near anti-phase conditions. After about 2.3\% difference in their frequencies, no definite relationship exists between the phases and the metronomes move away from a synchronized state.

\section{Conclusions}
\label{sec:VI}
To summarize, our paper studies both experimentally and numerically the synchronization of coupled self sustained oscillators, two non-identical metronomes in our system. Previous works in the topic have explored the in-phase synchronization in this system or anti-phase synchronization by adjusting a external parameter (base damping) \cite{Pantaleone, Wu, Ulrichs}. In contrast we have explored the transition of the system from an in-phase to anti-phase synchronized state by controlling two internal parameters, uncoupled frequency and internal damping. Our design of t experiments and the data acquisition system is  easily extendable to three or more oscillators.

It has been observed in the simulations that the two metronomes settle in a phase locked state when these parameters are mismatched, where the constant phase difference increases with an increase in the parameter mismatch, reaching near anti-phase synchronized conditions before moving away from synchronization with a further increase in the parameter mismatch. In the experiments it is observed that the metronomes move to a anti-phase synchronized state with a small mismatch in the parameters and complete in-phase synchronization is only obtained in a limited range.

Qualitatively, the effect of mismatching both parameters, internal damping or frequency is similar though the range upto which this mismatch can be sustained is quite different. In the simulations, the phase synchronization for damping mismatch occurs in the range of 0-26\% after which $\theta_1$ and $\theta_2$ becomes uncorrelated whereas this range is only 0 to about 2.4\% for frequency mismatch.

It might be speculated that the results from the original Huygens' experiment of coupled clocks, where the clocks were seen to only synchronize in antiphase condition \cite{Pikovsky} a certain contribution may also have come from this reason of parameter mismatch. It is possible that the clocks from the 17th century had a small difference in their damping parameters due to variability in their manufacturing process. Due to this mismatch in their damping parameter (even though their frequencies were probably tested more carefully and made to be near equal) the system moved away from synchronizing in complete in-phase condition and was only observed to synchronize in anti-phase (phase difference of 180 degrees) condition, which is what has been verified in our system.
\vspace{1cm}
\section{Acknowledgement}
 This work is supported by research grants from Industrial Research and Consultancy Center and Department of Physics, Indian Institute of Technology, Bombay and DST, Govt. of India. The authors thank Mr. Tanu Singla for useful comments and discussions.


\section{Appendix: Ellipse Fitting by Taubin method}
As described in Section \ref{sec:V} the $\theta_1$ vs. $\theta_2$ curve is not a smooth ellipse and there is a spread along each axis due to small vibrations around the limit cycle. Also different initial conditions leads to nearly the same ellipse.

To do this, we use a doubly optimal ellipse fitting routine for estimating planar curves determined by implicit equation. The method is also known as the Taubin method and the details are available in \cite{Taubin,Taubin2}.

The ellipse fitting routine returns the normalized vector A=$\{a,b,c,d,e,f\}$, such that:
\begin{equation}
 a\theta_1^2 + b\theta_1\theta_2 + c\theta_2^2 + d\theta_1 + e\theta_2 + f = 0
\end{equation}
Comparing with equation $(4.1)$ we have (Let N be the normalizing constant) :
\begin{eqnarray}
\frac{1}{A^2} = Na\nonumber\\
\frac{1}{B^2}= Nc \nonumber\\
\mbox{or,  } N\sqrt{ac} = \frac{1}{AB}
\end{eqnarray}

\begin{equation}
\frac{-2\cos\phi}{AB} = Nb
\end{equation}

Therefore putting Eq.(13) in the last equation we have:
\beq
 \phi = \cos^{-1} \frac{-b}{2\sqrt{ac}}
\eeq

Using this method, the constant phase difference of the two oscillators is found.


\end{document}